# Focusing properties of discrete RF quadrupole[*]

LI Zhi-Hui（李智慧）[1)], WANG Zhi-Jun （王志军）[2)]

1) The Key Laboratory of Radiation Physics and Technology of Ministry of Education, Institute of Nuclear Science and Technology, Sichuan University, Chengdu 610065, China

2) Institute of Modern Physics, CAS, Lanzhou, 730000, China

Abstract: The particle motion equation in the Radio Frequency (RF) quadrupole is derived. The motion equation shows that the general transform matrix of RF quadrupole with length less than or equal to $0.5\beta\lambda$ ($\beta$ is the relativistic velocity of particles and $\lambda$ is wavelength of radio frequency electromagnetic field) can describe the particle motion in arbitrary long RF quadrupole. By iterative integration, the general transform matrix of discrete RF quadrupole is derived from the motion equation. The transform matrix is in form of power series of focusing parameter B. It shows that for length less than $\beta\lambda$, the series up to 2nd order of B agrees well with the direct integration results for B up to 30, while for length less than $0.5\beta\lambda$, the series up to 1st order is a good approximation of the real solution for B less than 30 already. The formula of the transform matrix can be integrated into the linac or beam line design code to deal with the focusing of the discrete RF quadrupole.

Key words: RF quadrupole, Transform matrix, Focusing parameter, Iterative integration, Hybrid RFQ

# 1 Introduction

Inspired by the idea of separate accelerating and focusing, which was first implemented in the Interdigital H-type Drift Tube Linac (IH-DTL) structures [1], a new accelerating structure named Hybrid RF Quadrupole (H-RFQ) was proposed by P.N. Ostroumov in 2005 [2]. H-RFQ consists of an alternating series of drift tubes (DTLs) and radio frequency quadrupole (RFQ) sections, which are incorporated into one resonator. It has both advantages of high acceleration efficiency of IH-DTL structures and of efficient focusing of low energy ion beams of radio frequency (RF) quadrupole. Because of its superiority in accelerating low charge heavy ion beams, it has attracted attentions of some labs [3] since its invention.

One of the most important issues in designing the H-RFQ structure is to describe the focusing effect of the RF quadrupoles precisely. The focusing properties of RF quadrupole have been studied intensively during the development of RFQ accelerators in the 1970's, and one of the most important conclusions is that the transverse focusing strength of a unit cell is constant for uniform RF quadrupole structure. P.N. Ostroumov applied this to the discrete RF quadrupole and used a static magnetic quadrupole to replace the RF quadrupole [2]. He assumed that the RFQ sections consists of cells with length exactly equal to $\beta\lambda/2$, and the focusing effect of the RFQ cells is equivalent to a magnetic quadrupole of length $\beta\lambda/2$ if

$$R_{0i} = \sqrt{\frac{2U_1}{\pi\beta c G_m}}, \tag{1}$$

where $U_1$ is the electric potential on the RFQ electrodes and $G_m$ is the equivalent magnetic gradient. Clearly it takes the RFQ cell as a static electric quadrupole with potential equal to the average potential within half RF period

$$V_{eq} = \frac{\int_{-\beta\lambda/4}^{\beta\lambda/4} U_1 \cos kz\, dz}{\int_{-\beta\lambda/4}^{\beta\lambda/4} dz} = \frac{2U_1}{\pi}. \tag{2}$$

---
[*] Supported by the National Natural Science Foundation of China (11375122, 11511140277) and by the Strategic Priority Research Program of the Chinese Academy of Sciences，Grant No. XDA03020705）

1) E-mail:lizhihui@scu.edu.cn

The magnetic quadrupole with equal focusing strength has magnetic gradient

$$G_m = \frac{2U_1}{\pi\beta c}\frac{1}{R_{0i}^2}. \tag{3}$$

If we investigate the assumptions made in reference [1] carefully, we will find that they do not stand on a solid foundation. Firstly, the length of the RFQ cell is not necessarily equal to $\beta\lambda/2$ if the accelerating gap length is taken into account. For IH structures, the gap length is normally about 1/3 to 1/2 of the period length, so the RFQ cell length is only 2/3 or 1/2 of $\beta\lambda/2$ if the synchronous phases of gaps on the both sides of RFQ section are equal. Furthermore, in order to avoid the RF sparking, the gap lengths between drift tube and RF quadrupole section are even bigger, and this makes the length of RFQ cell even shorter. Secondly, for a discrete RFQ cell, the focusing strength depends on the RF phase when particle get into the RFQ section, so it cannot be equivalent to a static magnetic quadrupole. In this paper, the general solution of RF electric quadrupole in matrix form will be derived, which can be integrated into the linac design code.

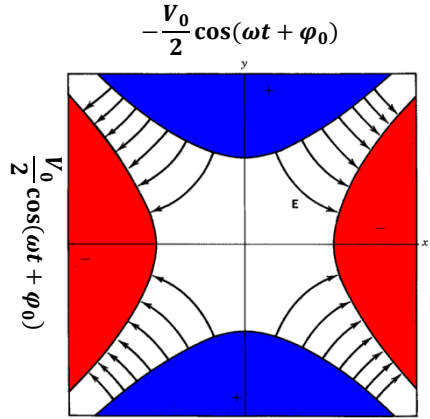

Fig.1 the cross section of RF quadrupole

## 2 Motion equation and transform matrix

The cross section of RF electric quadrupole is shown in Fig.1. We define x and y axes as shown in Fig.1, and z axis is directed outside of the paper from origin of xy axes. If we choose the time when particles get into the RFQ as zero of time, since the electrodes are uniform in z and electric field line is perpendicular to the z direction, the particle velocity along z direction is constant, so we have

$$\omega t = kz, \tag{4}$$

where ω is the angle frequency of the RF field, and $k$ is the wave number and is equal to $2\pi/\beta\lambda$. The electric potential of electrodes in xz plane is

$$V_{RFQ} = \frac{V_0}{2}\cos(kz + \varphi_0), \tag{5}$$

where $\varphi_0$ is the initial RF phase when particles get into the RFQ. The potential of the electrodes in yz plane is $-V_{RFQ}$. Since the motions in xz plane and yz plane are independent, we just consider the motion in xz plane in the following. The motion equation can be derived from Newton's law. If taken the distance along z axis as independent variable, then the motion equation in xz plane is

$$\frac{d^2x}{dz^2} + \frac{1}{\gamma mc^2\beta^2}\frac{qV_0}{a^2}\cos(kz + \varphi_0)\,x = 0. \tag{6}$$

where γ is the relativistic parameter, q is the charge of the particle and a is the aperture radius of the quadrupole. For that in yz plane, we just need to replace $V_0$ and x with $-V_0$ and y, respectively. This equation is the famous Mathieu equation [4] and has been studied intensively. It has four kinds of solutions and they are very complicated in form. In order to get a set of solutions which can be expressed in the form of transfer matrix, so that it can be easily integrated in the linac design code, we need to rewrite equation (6) in another form. The well-known transverse focusing parameter B of RFQ accelerator is defined as [5]

$$B = \frac{qXV_0}{\gamma mc^2}\frac{\lambda^2}{a^2} \tag{7}$$

where $X$ is the focusing parameter. In our case, there is no modulation of the electrodes, so $X=1$. The motion equation can be written as two first order equations

$$\frac{dx}{dz} = x' \tag{8}$$

$$\frac{dx'}{dz} = -\frac{B}{\beta^2\lambda^2}\cos(kz + \varphi_0)\,x \tag{9}$$

Suppose the transform matrix of equation (9) is R, then the coordinates at z are related with initial coordinates by R as,

$$\begin{pmatrix}x(z)\\x'(z)\end{pmatrix} = R(z,\varphi_0,B)\begin{pmatrix}x_0\\x'_0\end{pmatrix} \tag{10}$$

Where $R$ is a matrix of function of $z$, initial phase $\varphi_0$ and focusing parameter $B$. Defining $R(0.5\beta\lambda,\varphi_0,B)$ as the transform matrix of quadrupole with length $0.5\beta\lambda$ and initial phase $\varphi_0$, we have the following relation from the motion equation,

$$\begin{pmatrix}x(z + 0.5\beta\lambda,\varphi_0,B)\\x'(z + 0.5\beta\lambda,\varphi_0,B)\end{pmatrix} = R(z,\varphi_0 + \pi,B)R(0.5\beta\lambda,\varphi_0,B)\begin{pmatrix}x_0\\x'_0\end{pmatrix} \tag{11}$$

It indicates that once we obtain the general transform matrix of RF quadrupole with length less than or equal to $0.5\beta\lambda$, then with equation (11) we have the transform matrix of RF quadrupole with arbitrary length.

Suppose the initial conditions of the particle are $(x_0, x'_0)$. For the first step, we suppose that the velocity of particle is kept constant and is $x'_0$, i.e.,

$$x' = x'_0 = a_{21}^{(0)}x_0 + a_{22}^{(0)}x'_0 \tag{12}$$

Taking equation (12) into equation (8) we have that

$$x = x_0 + x_0'z = a_{11}^{(0)}x_0 + a_{12}^{(0)}x_0' \tag{13}$$

Defining the transfer matrix of zero order on $B$ as

$$R^0 = \begin{pmatrix} a_{11}^{(0)} & a_{12}^{(0)} \\ a_{21}^{(0)} & a_{22}^{(0)} \end{pmatrix} = \begin{pmatrix} 1 & z \\ 0 & 1 \end{pmatrix} \tag{14}$$

This is just the same as the transfer matrix of a drift space with length z. Now we take equation (13) as the first approximation of the transverse location of the particle and take it into equation (9) and integrate it, we obtain x' in the first order approximation on B, that is

$$x' = x_0' - \frac{B}{\beta^2\lambda^2}\int_0^z \cos(kz+\varphi_0)(x_0 + x_0'z)dz = \left[a_{21}^{(0)} + a_{21}^{(1)}\right]x_0 + \left[a_{22}^{(0)} + a_{22}^{(1)}\right]x_0'. \tag{15}$$

Taking equation (15) into equation (8), we obtain x in the first order of approximation on B.

$$x = x_0 + \int_0^z(\left[a_{21}^{(0)} + a_{21}^{(1)}\right]x_0 + \left[a_{22}^{(0)} + a_{22}^{(1)}\right]x_0')dz = \left[a_{11}^{(0)} + a_{11}^{(1)}\right]x_0 + \left[a_{12}^{(0)} + a_{12}^{(1)}\right]x_0' \tag{16}$$

where,

$$a_{11}^{(1)} = \frac{B}{(2\pi)^2}[\cos\varphi_0 - \cos(kz+\varphi_0) - kz\sin\varphi_0] \tag{17-1}$$

$$a_{12}^{(1)} = -\frac{B}{(2\pi)^2}\left[z\cos\varphi_0 + z\cos(kz+\varphi_0) + \frac{2}{k}\sin\varphi_0 - \frac{2}{k}\sin(kz+\varphi_0)\right] \tag{17-2}$$

$$a_{21}^{(1)} = -\frac{B}{(2\pi)^2}k[\sin\varphi_0 - \sin(kz+\varphi_0)] \tag{17-3}$$

$$a_{22}^{(1)} = -\frac{B}{(2\pi)^2}[\cos\varphi_0 - \cos(kz+\varphi_0) - kz\sin(kz+\varphi_0)] \tag{17-4}$$

If we investigate the procedure above, we can find it is very similar to the well-known "drift-kick" numerical method in dealing with such problems [6], i.e., dividing the whole RFQ section into small sections, and for each section, the field can be looked as constant, as the section number increased, the solution will approach to the real solution with very high precision. If we repeat the procedure above, we obtain the results on the second order approximation on $B$ as,

$$x' = \left[a_{21}^{(0)} + a_{21}^{(1)} + a_{21}^{(2)}\right]x_0 + \left[a_{22}^{(0)} + a_{22}^{(1)} + a_{22}^{(2)}\right]x_0' \tag{18}$$

$$x = \left[a_{11}^{(0)} + a_{11}^{(1)} + a_{11}^{(2)}\right]x_0 + \left[a_{12}^{(0)} + a_{12}^{(1)} + a_{12}^{(2)}\right]x_0' \tag{19}$$

where

$$a_{11}^{(2)} = \frac{B^2}{8(2\pi)^4}[12 - 2k^2z^2 - 5\cos2\varphi_0 - 12\cos(kz) + \cos(2kz+2\varphi_0) + 4\cos(kz+2\varphi_0) +$$
$$2kz\sin2\varphi_0 - 4kz\sin(kz) + 4kz\sin(kz+2\varphi_0)] \tag{20-1}$$

$$a_{12}^{(2)} = \frac{B^2}{24(2\pi)^4}\Big[36z - 2k^2z^3 + 3z\cos2\varphi_0 + 12z\cos kz + 3z\cos(2kz+2\varphi_0) + 12z\cos(kz+$$
$$2\varphi_0) + \frac{9}{k}\sin2\varphi_0 - \frac{48}{k}\sin(kz) - \frac{9}{k}\sin(2kz+2\varphi_0)\Big] \tag{20-2}$$

$$a_{21}^{(2)} = -\frac{B^2}{4(2\pi)^4}k[2kz + 2kz\cos(kz) - 2kz\cos(kz+2\varphi_0) - \sin2\varphi_0 - 4\sin(kz) + \sin(2kz+$$
$$2\varphi_0)] \tag{20-3}$$

$$a_{22}^{(2)} = \frac{B^2}{8(2\pi)^4}[12 - 2k^2z^2 + \cos2\varphi_0 - 12\cos(kz) - 5\cos(2kz+2\varphi_0) + 4\cos(kz+2\varphi_0) -$$

$$4kzsin(kz) - 2kzsin(2kz + 2\varphi_0) - 4kzsin(kz + 2\varphi_0)] \tag{20-4}$$

In general, we can find the transfer matrix of RF electric quadrupole is expressed as

$$R = R^{(0)} + R^{(1)} + R^{(2)} + \cdots + R^{(i)} + \cdots, \tag{21}$$

where $R^{(i)}$ is proportional to the i'th power of $B/4\pi^2$. Usually, $B$ is less than 10 for RFQ accelerators.

## 3 Discussion

From preceding section we know that the transform matrix is a power series of $B/4\pi^2$, and the number of terms increases with the increase of the power number. It is hard to get its general convergent property in theory. According to the definition of transform matrix, if the initial coordinates of the particle are (1, 0), we have the particle coordinates at z as

$$\begin{pmatrix} x(z) \\ x'(z) \end{pmatrix} = R \begin{pmatrix} 1 \\ 0 \end{pmatrix} = \begin{pmatrix} a_{11} \\ a_{21} \end{pmatrix}. \tag{22}$$

If the initial coordinates of the particle are (0,1), then the particle coordinates at z are

$$\begin{pmatrix} x(z) \\ x'(z) \end{pmatrix} = R \begin{pmatrix} 0 \\ 1 \end{pmatrix} = \begin{pmatrix} a_{12} \\ a_{22} \end{pmatrix}. \tag{23}$$

That means the elements of transform matrix are the solutions of the motion equations with special initial conditions. In order to verify the validation condition and its precision, we calculated the transform matrix by integrating the motion equations (8) and (9) directly with initial conditions (1, 0) and (0, 1), and compared the results with those calculated from the formulas we get in the last section to see where the series can be truncated and its precision.

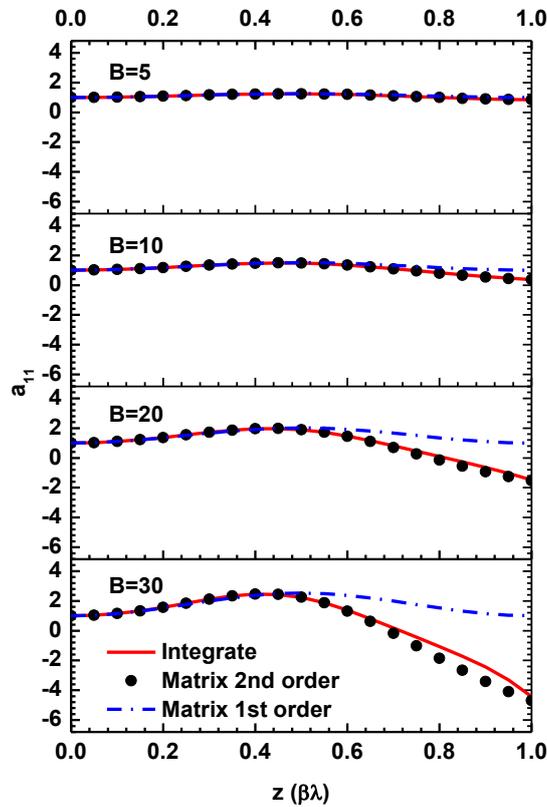

Fig.2 $a_{11}$ as function of z (the initial phase is 0 degree. The solid red lines: integration of the

motion equation; dash-dot blue lines: matrix in 1st order; black dot: matrix in 2nd order)

Fig.2 shows $a_{11}$ calculated by motion equation integration and transform matrix formulas in first and second order approximation, where the initial phase is 0 degree. We can see, for B=5, the results calculated in three ways agree well for z from zero to $\beta\lambda$; as B increase, the deviation between the 1st order formula and the other two curves becomes larger when z is greater than $0.5\beta\lambda$, while the 2nd order formula and the integration results agree well except for B=30 case. It indicates that the 2nd order formula can approximate the motion equation well for B up to 30 and z less than $\beta\lambda$, and for $z<0.5\beta\lambda$ the 1st order formula is already accurate enough.

Fig. 3 shows the four transform matrix elements calculated by three different ways, where B is set to 10, and the RF phase when particles get into the RF electric quadrupole is -90 degree. We can see the second order matrix elements agree with the integration results very well for z less than $\beta\lambda$. For $z<0.5\ \beta\lambda$, the 1st order formula is a good approximation.

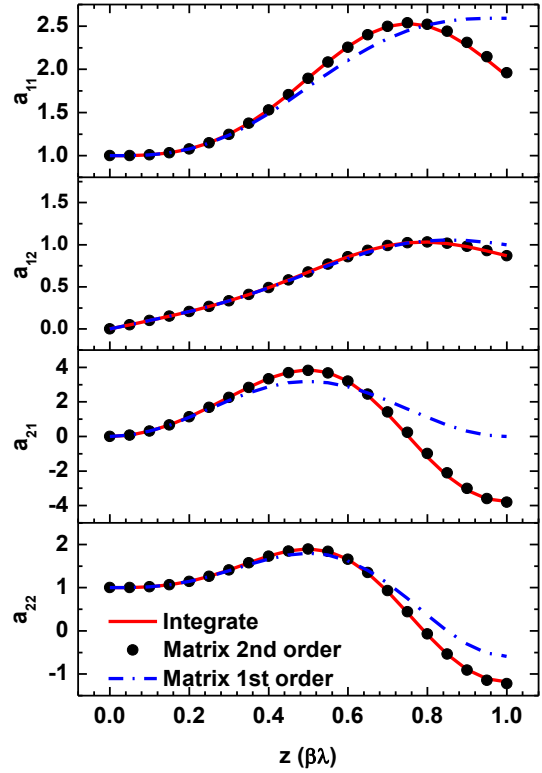

Fig.3 the results of transform matrix elements (B=10 and initial phase is -90 degree)

From equation (11), we have concluded that the transform matrix which is valid for $z \leq 0.5\beta\lambda$ is just needed. From the discussions above we find that the maximum deviation happens at $0.5\beta\lambda$. Since $R(0.5\beta\lambda, \varphi_0, B)$ may be used several time for long quadrupole, it is better to use the 2nd order form for $R(0.5\beta\lambda, \varphi_0, B)$. However for $R(z, \varphi_0, B)$ with $z<0.5\beta\lambda$, the 1st order form can be applied for simplity. Let $kz=\pi$ and take it into equations (17) and (20), we obtain the formula of $R(0.5\beta\lambda, \varphi_0, B)$ as following,

$$R(0.5\beta\lambda, \varphi_0, B) = \begin{pmatrix} a_{\pi,11} & a_{\pi,12} \\ a_{\pi,21} & a_{\pi,22} \end{pmatrix} \tag{24}$$

where

$$a_{\pi,11} = 1 + \frac{B}{(2\pi)^2}(2\cos\varphi_0 - \pi\sin\varphi_0) + \frac{B^2}{(2\pi)^4}\left[3 - \frac{\pi^2}{4} - \cos2\varphi_0 - \frac{\pi}{4}\sin2\varphi_0\right] \qquad (25\text{-}1)$$

$$a_{\pi,12} = \frac{1}{2}\beta\lambda\left[1 - \frac{B}{(2\pi)^2}\frac{4}{\pi}\sin\varphi_0 + \frac{B^2}{(2\pi)^4}\left(1 - \frac{\pi^2}{12} - \frac{1}{4}\cos2\varphi_0\right)\right] \qquad (25\text{-}2)$$

$$a_{\pi,21} = -\frac{2\pi}{\beta\lambda}\left[\frac{B}{(2\pi)^2}2\sin\varphi_0 + \frac{B^2}{(2\pi)^4}\frac{\pi}{2}\cos2\varphi_0\right] \qquad (25\text{-}3)$$

$$a_{\pi,22} = 1 - \frac{B}{(2\pi)^2}(2\cos\varphi_0 + \pi\sin\varphi_0) + \frac{B^2}{(2\pi)^4}\left[3 - \frac{\pi^2}{4} - \cos2\varphi_0 + \frac{\pi}{4}\sin2\varphi_0\right] \qquad (25\text{-}4)$$

## 4 Conclusions

The discrete RF quadrupole can not be equivalent to the static quadrupole, and its focusing strength is the function of RF phase when particles get into the quadrupole, so the transverse motion is coupled with the longitudinal motion for RF accelerators which apply discrete RF quadrupole as transverse focusing elements. The transform matrix of RF quadrupole is a power series of focusing parameter B. When the quadrupole length is less than βλ and B is less than 30, the series up to 2nd order of B can precisely describe the focusing properties of the RF quadrupole, while if the quadrupole length is less than 0.5βλ and B is less than 30, the series up to 1st order of B agrees well with the real solution of the motion equation. With 2nd order transform matrix of RF quadrupole with length 0.5βλ and 1st order transform matrix of quadrupole with length less than 0.5βλ, the particle motion in an arbitrary long RF quadrupole can be described.